\documentclass[10pt, dvipsnames]{wlscirep}
\usepackage[utf8]{inputenc}
\usepackage[T1]{fontenc}

\title{\fontsize{19}{19}\selectfont Socio-economic disparities and COVID-19 in the USA}
\author[1,2,$\dag$]{Ayan Paul}
\author[1,$\dag$]{Philipp Englert}
\author[3,$\dag$]{Melinda Varga}

\affil[1]{DESY, Notkestra{\ss}e 85, D-22607 Hamburg, Germany}
\affil[2]{Institut f\"ur Physik, Humboldt-Universit\"at zu Berlin, D-12489 Berlin, Germany}
\affil[3]{2400 Elliot Ave., Seattle WA 98121, USA}
\affil[$\dag$]{e-mail: \href{mailto:ayan.paul@desy.de}{ayan.paul@desy.de}, \href{phillip.englert@desy.de}{philipp.englert@desy.de}, \href{mailto:mvarga@alumni.nd.edu}{mvarga@alumni.nd.edu}}


\begin{abstract}
COVID-19 is not a universal killer. We study the spread of COVID-19 at the county level for the United States up until the 15$^{th}$ of August, 2020. We show that the prevalence of the disease and the death rate are correlated with the local socio-economic conditions often going beyond local population density distributions, especially in rural areas. We correlate the COVID-19 prevalence and death rate with data from the US Census Bureau and point out how the spreading patterns of the disease show asymmetries in urban and rural areas separately and are preferentially affecting the counties where a large fraction of the population is non-white. Our findings can be used for more targeted policy building and deployment of resources for future occurrence of a pandemic due to SARS-CoV-2. Our methodology, based on interpretable machine learning and game theory, can be extended to study the spread of other diseases. 
\end{abstract}

\begin{document}
\flushbottom
\maketitle

\thispagestyle{fancy}
\rhead{DESY 20-134, HU-EP-20/20}

\section{Introduction}
\label{sec:intro}

The rapid spreading of the novel SARS-CoV-2 virus left very little options for carefully curating mitigation measures. This has led to public health and disease mitigation policies being determined at the national or state level that often failed to take into account variability at a smaller geographical scale. It is well understood that a disease that is spread by proximity and contact will have a larger impact in more densely populated regions and areas with higher mobility. However, the footprint of socio-economic factors on the spread of the disease cannot be discounted in areas where larger population density alone does not drive the spread of the disease. Local socio-economic conditions add a significant amount of variability in the containment of any disease spread. These conditions, along with ethnicity, impact HIV prevalence and its delayed diagnosis~\cite{Ransome2016}. Studies on the spread of pathogens like influenza, MERS-CoV, SARS-CoV, Ebola etc. indicate that socio-economic conditions should be taken into account while building policies for resource deployment during the onset of a pandemic~\cite{Farmer1996,10.1371/journal.pone.0012763,Quinn2014}. Even beyond the variances in socio-economic conditions, ethnicity is a contributing factor~\cite{NAP12875,10.1001/jama.1990.03440170066038,PMID:11110355,PMID:9386949} for disparities in healthcare and more so for health conditions that require intensive and prolonged care~\cite{doi:10.1161/STR.0b013e3182213e24}. These disparities can also incorporate stereotyping of race and ethnicity within the health care system~\cite{1999-03648-011}. 

The communities of racial minorities, both black and Hispanic, have been disproportionately affected by COVID-19~\cite{MILLETT2020,10.1001/jama.2020.6548,doi:10.1056/NEJMp2021971,10.15585/mmwr.mm6933e1,10.1001/jama.2020.11374,DIMAGGIO20207,MILLETT202037,doi:10.1080/13557858.2020.1853067,Pareek2020,Laurencin2020} from the early stages of the pandemic. A recent study~\cite{Goyale2020009951} of 1000 children tested for SARS-CoV-2 showed that minority children have a higher odds of testing positive with non-Hispanic blacks having an adjusted odds of infection of 2.3 and Hispanics having an adjusted odds of infection of 6.3 over non-Hispanic whites. Socio-economic differences can also surface in the form of difficulties of imposing shelter-in-place~\cite{UchicagoWP,Weill19658} leading to failures in mitigation methods and allowing for the spread of the virus. A study of the COVID-19 outbreak in Wuhan in January and February focused on the effects of socio-economic factors on the transmission patterns of the disease~\cite{Qiu2020}. This study shows that cities with better socio-economic resources allowing for better infrastructure and response were better able to contain the spread of the disease. An extensive study of the effects of a large set of socio-economic conditions that affect the transmission of COVID-19 has recently been completed~\cite{2020arXiv200407947S}. This work studies the importance of various socio-economic metrics in determining the disease spread and their correlations. The primary difference between their work and ours is that we study variation in the transmission pattern of COVID-19 within one country at the county level as opposed to comparing various countries. 

In an attempt to keep our study very general we look at a broad spectrum of socio-economic data collected from the 5-years American Census Survey 2018 of the United States Census Bureau. This not only leads to a better understanding of the impact that socio-economic conditions have on disease propagation, but also gives us a prescriptive method of tying together socio-economic conditions and the pattern of spread of COVID-19. This insight can possibly lead to the building of better policies, even at the county level, to address the ongoing and any future pandemic caused by COVID-19. We believe that the currently available data on COVID-19 prevalence has reached a stage of maturity having evolved through the stages of unchecked and mitigated spread. This adds confidence to the conclusion that we draw in this work. To understand the complex interdependence between socio-economic data and disease prevalence along with population demographics, we bring to this work the application of boosted decision trees to survey the data and learn the embedded correlations. In addition we borrow a method from game theory, called Shapley values, to understand the relative importance of various socio-economic markers and population density in determining COVID-19 prevalence and death rate. In what follows we will give a brief description of how we curate the data and then present some details of our novel analysis methods. In our analysis we will focus on some regions to highlight the features that are atypical in different parts of the country and would be lost in aggregation and then take a look at USA as a nation. We conclude with a summary of our results.

\section{Data sources and curation}
\label{sec:data}

\begin{figure}
    \centering
    \includegraphics[trim=30 0 20 30,clip,width=0.43\textwidth]{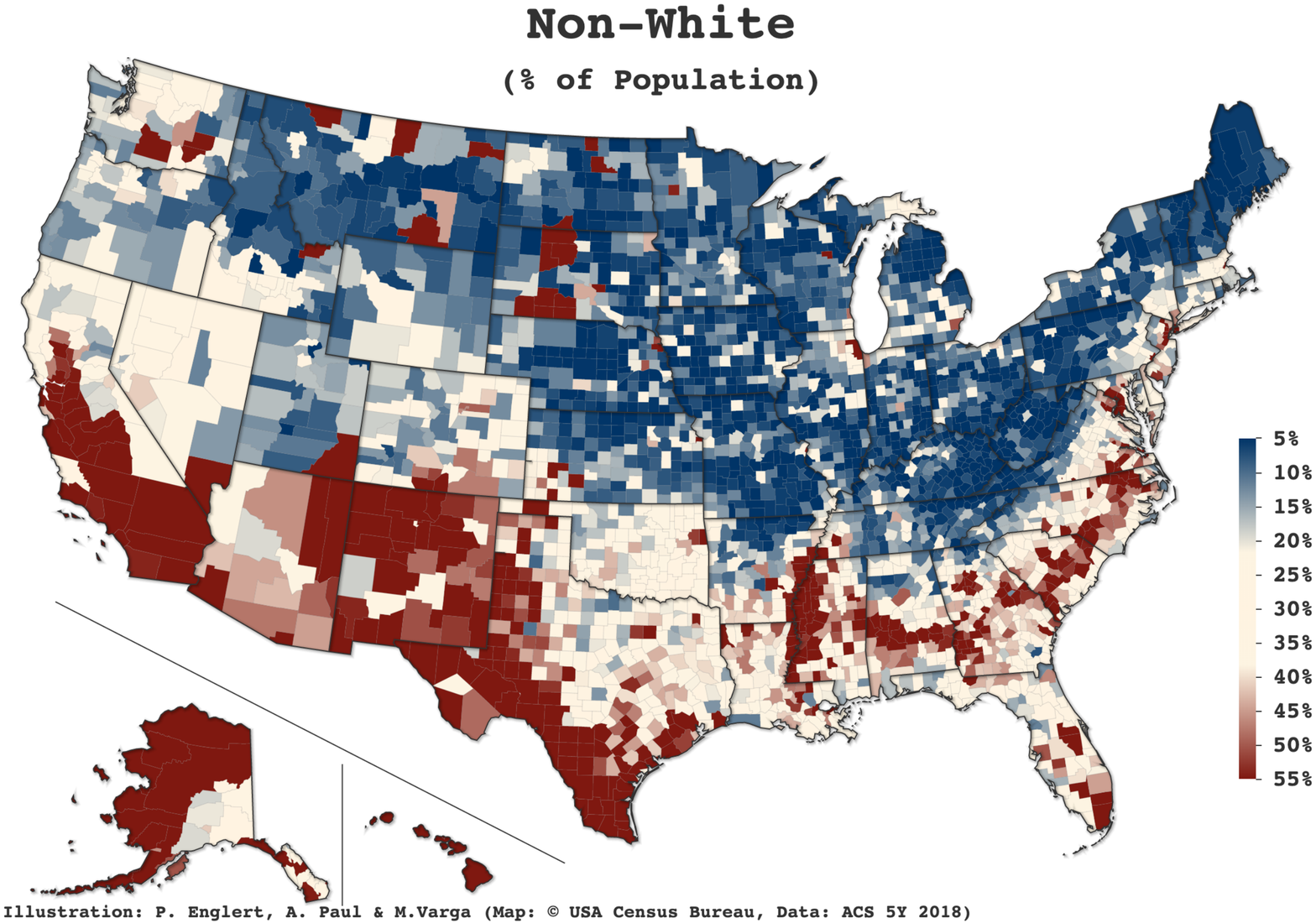}
    \includegraphics[trim=40 0 20 30,clip,width=0.43\textwidth]{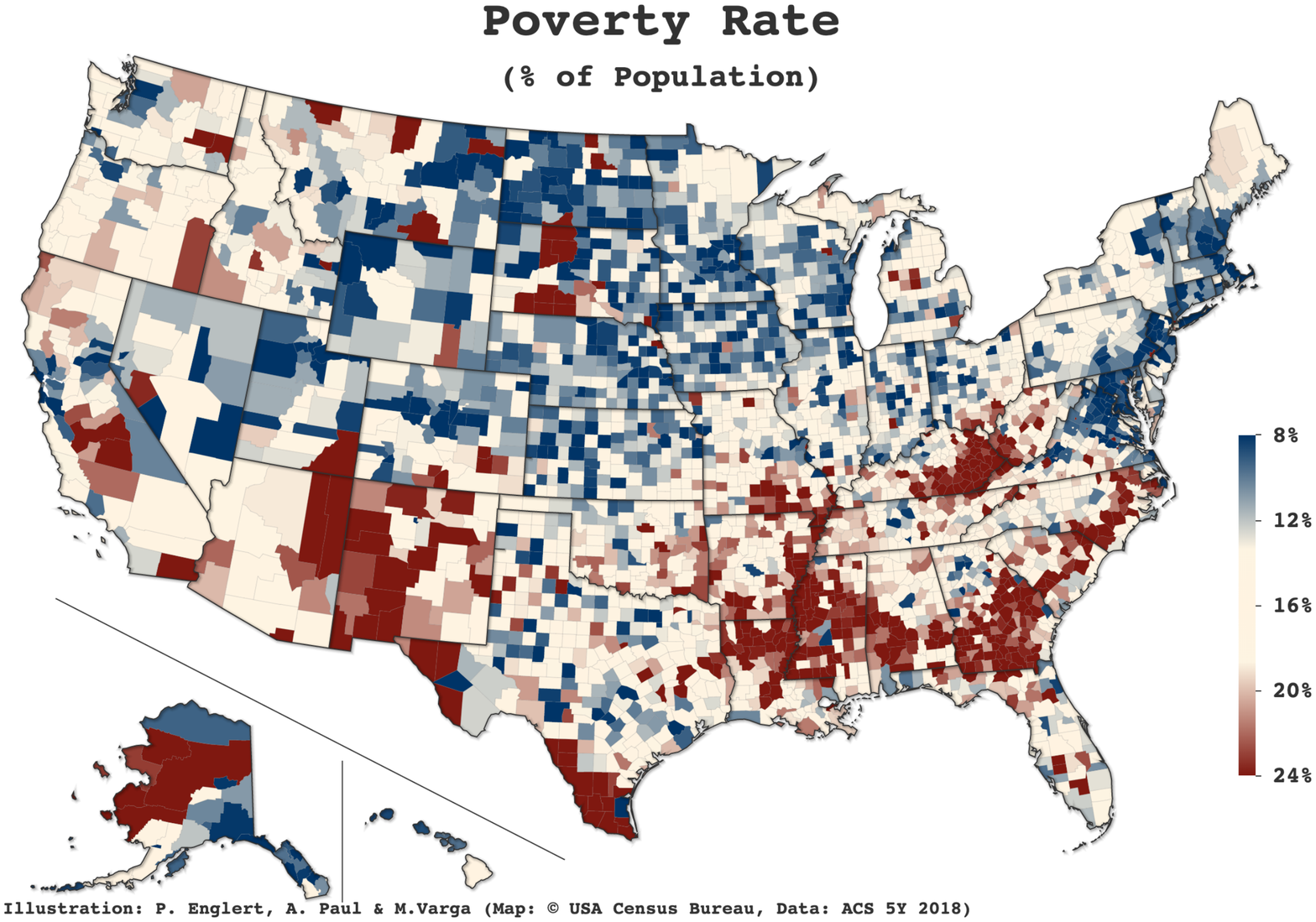}\\
    \includegraphics[trim=40 0 20 30,clip,width=0.43\textwidth]{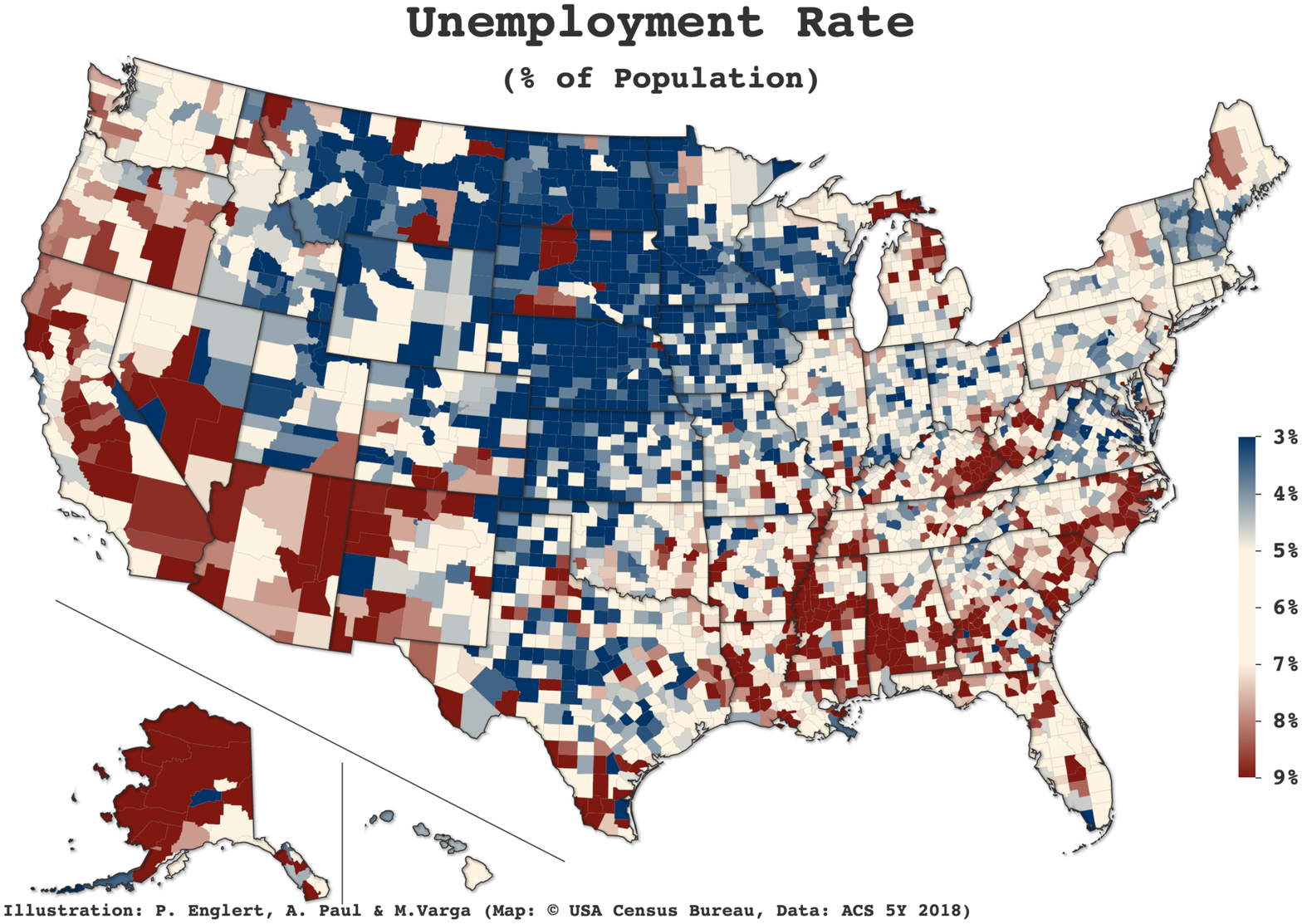}
    \includegraphics[trim=40 0 20 30,clip,width=0.43\textwidth]{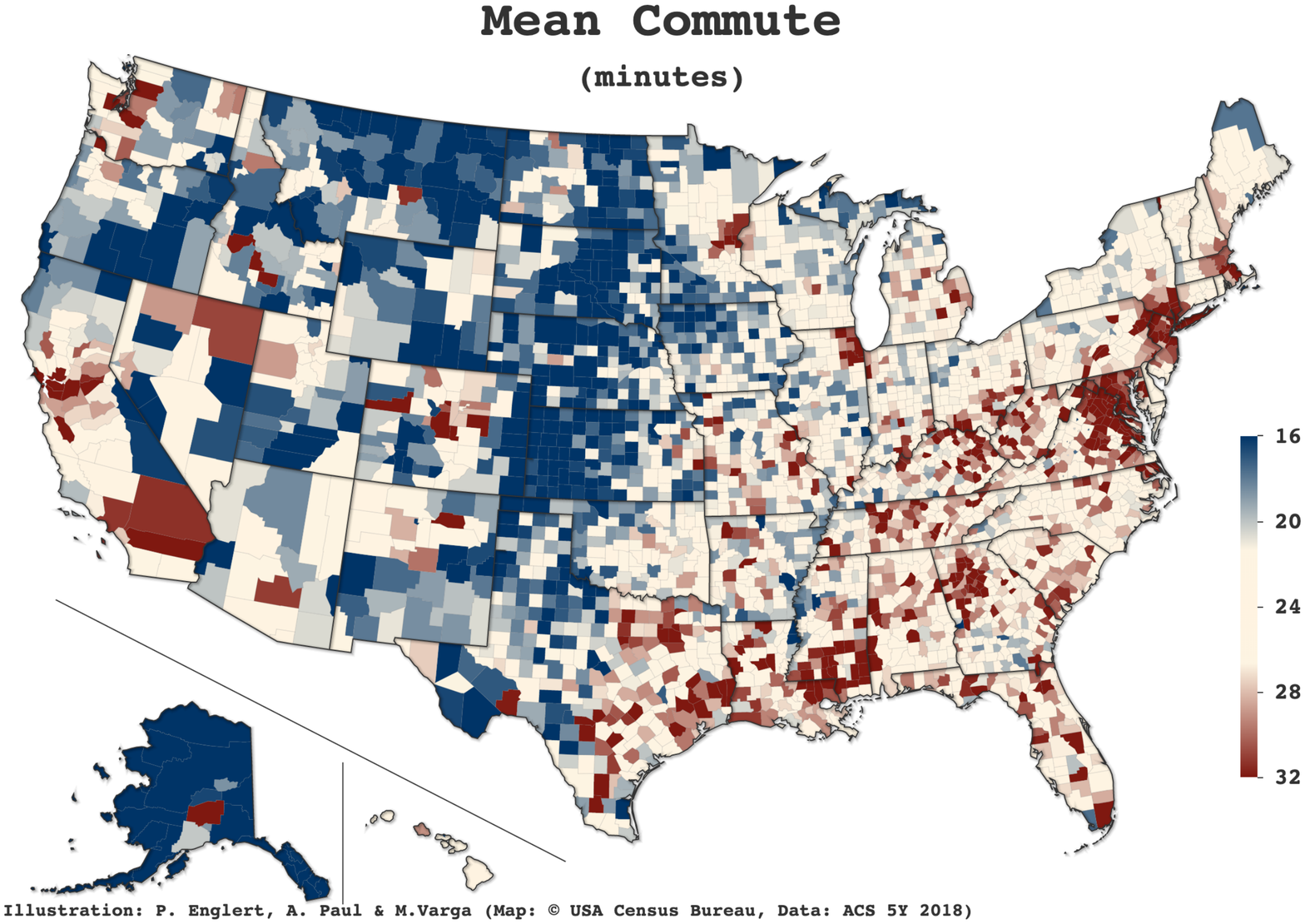}\\
    \includegraphics[trim=40 0 20 30,clip,width=0.43\textwidth]{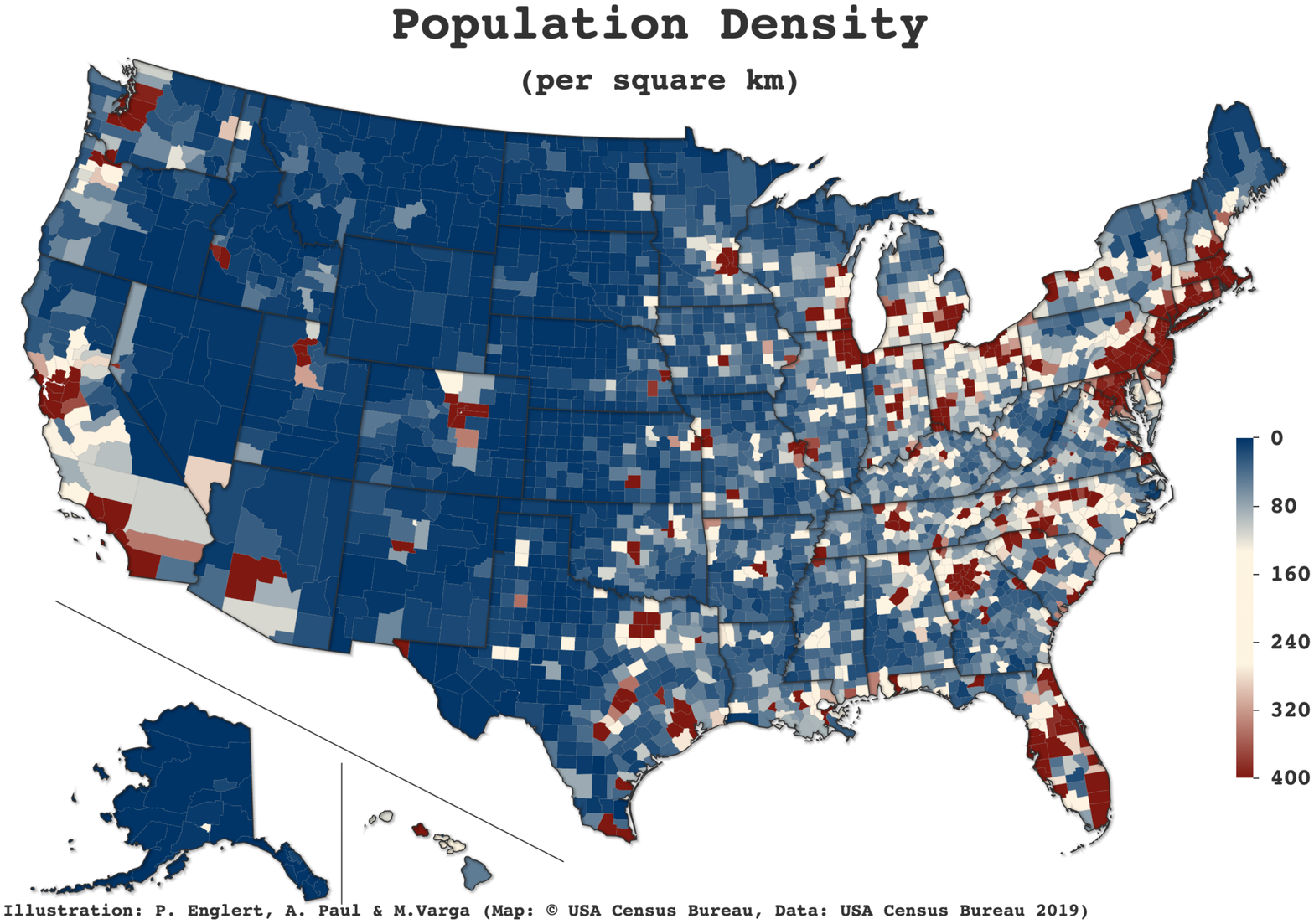}
    \includegraphics[trim=40 0 20 30,clip,width=0.43\textwidth]{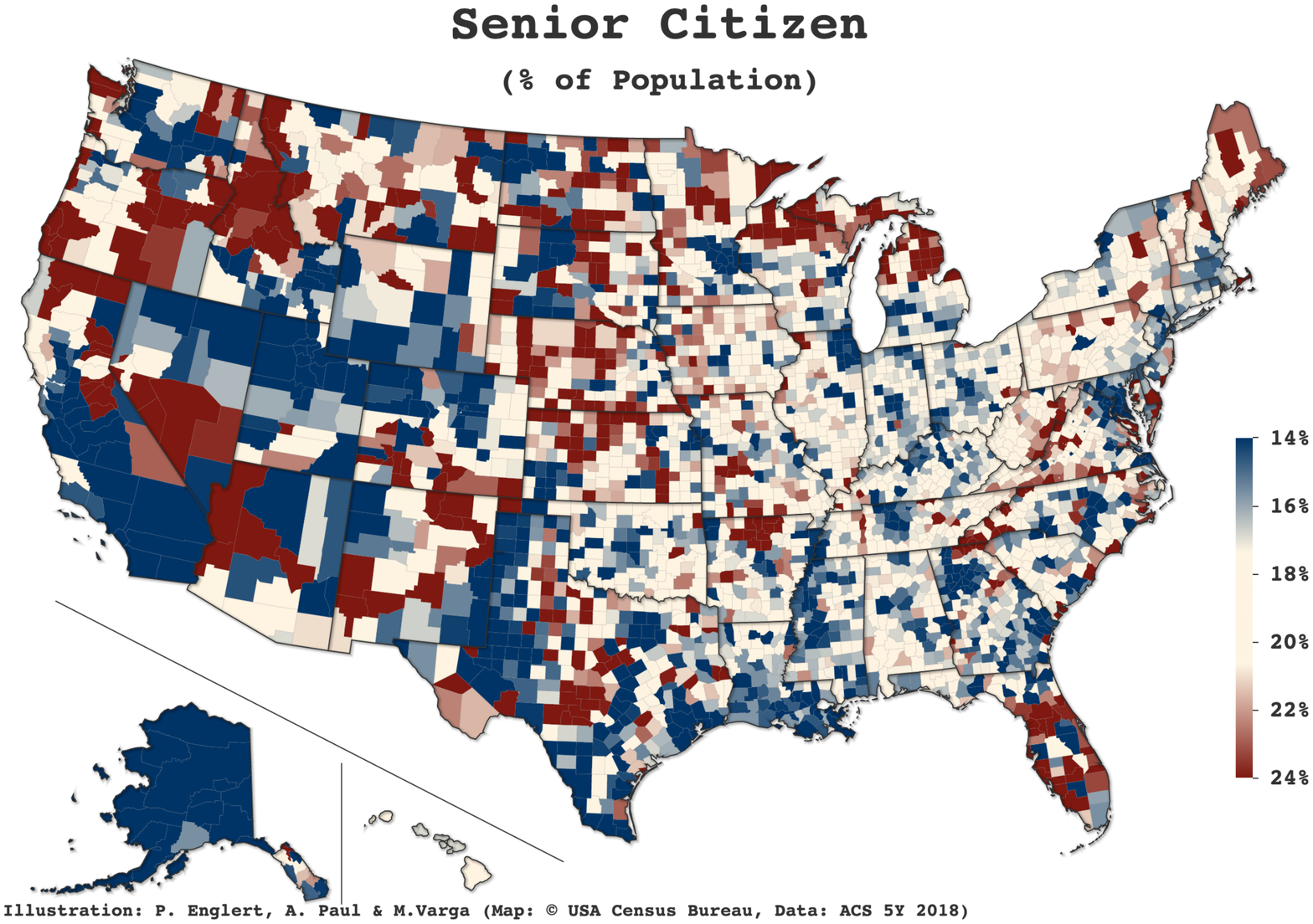}\\
    \includegraphics[trim=40 0 20 30,clip,width=0.43\textwidth]{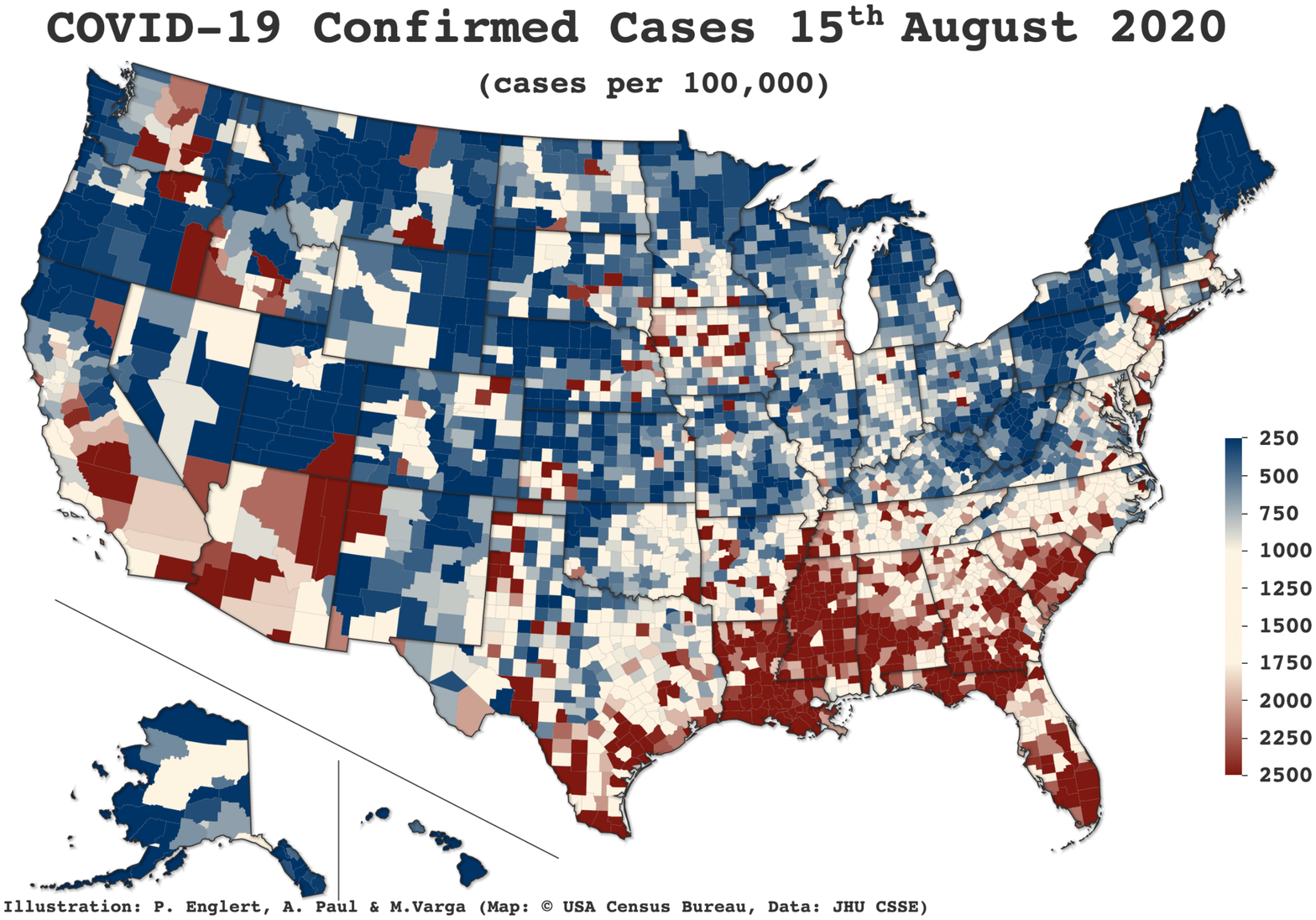}
    \includegraphics[trim=40 0 20 30,clip,width=0.43\textwidth]{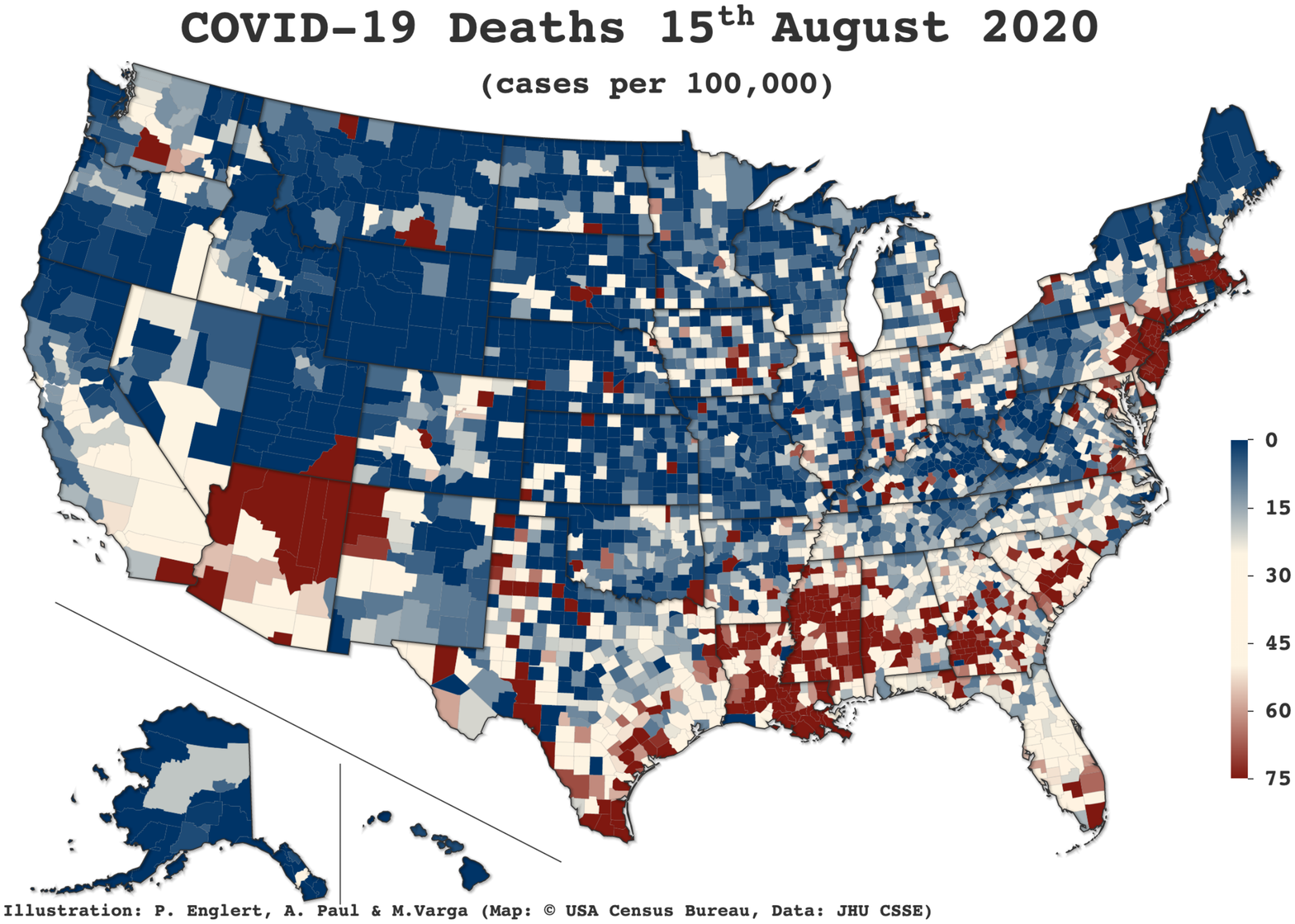}
    \caption{\it Distribution of some socio-economic metrics at the county level along with population density distribution. The two bottom panels show the COVID-19 confirmed cases and deaths per 100,000 individuals in each county respectively up until 15$^{th}$ of August 2020. The lower and upper bounds are set by 10 percentile and 90 percentile of the distributions respectively. Plots made with the Highcharts Maps JavaScript library from \href{https://www.highcharts.com/}{Highcharts.com} with a CC BY-NC 3.0 license.}
    \label{fig:maps_US}
\end{figure}

We collected the data used in this work from three primary sources:
\begin{itemize}
    \item The 2018 American Community Survey 5-years supplemental update to the 2011 Census found in the USA Census Bureau database for constructing the socio-economic metrics.
    \item The population density data that reflects the 2019 estimates of the US Census Bureau.
    \item Data on COVID-19 prevalence and death rate is obtained from the Johns Hopkins University, Center for Systems Science and Engineering database~\cite{Dong:2020ka} through their \href{https://github.com/CSSEGISandData/COVID-19}{GitHub repository}. We use the data ending 15$^{th}$ of August, 2020. 
\end{itemize}

Population density and mobility within the population are two factors that can be, a priori, deemed to be important for the spread of COVID-19. Since we are dealing with populations and not individuals, mobility has to be indirectly probed. One of the measures that probe the mobility within the population is any aggregate measure of the age of the population. For example, a population with a lower median age will, on an average, have higher mobility than a population with a much higher median age. Considering the fact that COVID-19 tends to affect the elderly preferentially, the fraction of a population that fall into the senior citizen category (over 65 years of age) is a good measure of quantifying both the mobility and the degree to which age plays a role in determining the spread of COVID-19.

To study the relevance of economic conditions in the spread of COVID-19, we focus on the metrics that quantify income, poverty, the employed fraction of the population and the unemployment rate. The latter two are not fully correlated since they add up to the fraction of the population that is employable which varies from county to county. To further probe the effects of mobility we chose to include the mean commute time in any county along with the fraction of the population that use a transit system. It is also known that certain professions put individuals at higher risk since they bring them at greater exposure to a larger number of people. This includes individuals employed in the service industry, construction, delivery, labor etc. Hence, we use a metric that quantifies the fraction of the population that work in these industries. Lastly, we include the fraction of individuals in a county that do not have health insurance as a metric to quantify how it affects the spread of the disease. The list of socio-economic metrics for each county that were used in this work are as follows:
\begin{itemize}
    \item {\bf Population Density}: The population density data taken from 2019.
    \item {\bf Non-White}: The fraction of non-white population in any county including Hispanics and Latinos.
    \item {\bf Income}: The income per capita as defined by the US Census Bureau.
    \item {\bf Poverty}: The fraction of the population deemed as being below the poverty line.
    \item {\bf Unemployment}: The unemployment rate as defined by the US Census Bureau.
    \item {\bf Uninsured}: The fraction of the population that does not have health insurance.
    \item {\bf Employed}: The fraction of the population that is employed.
    \item {\bf Labour}: The fraction of the population working in construction, service, delivery or production.
    \item {\bf Transit}: The fraction of the population who take the public transportation system or carpool excluding those who drive or work from home.
    \item {\bf Mean Commute}: The mean commute distance for a person living in a county in minutes.
    \item {\bf Senior Citizen}: The fraction of the population that is above 65 years of age.
\end{itemize}
For the data on COVID-19 prevalence we look at the total number of confirmed cases and deaths up until the 15$^{th}$ of August 2020 and define the following rates:
\begin{itemize}
    \item {\bf Confirmed Case Rate:} The total number of confirmed cases per 100,000 individuals in any county.
    \item {\bf Death Rate:} The total number of deaths per 100,000 individuals in any county.
    \item {\bf Fatality Rate:} This is the naive fatality rate and we do not adjust it for the delay between case confirmation and death. We have checked that this does not change the results of our analysis if we account for a delay of 10 days. The rate is defined as the total number of deaths over the total number of confirmed cases in any county. We will see that the naive fatality rate is mostly uncorrelated with the socio-economic metrics and we will not focus on it. The naive fatality rate represents the fraction of infected people that died once infected with COVID-19 and this seems to be mostly universal and independent of the socio-economic metrics and population density.
\end{itemize}

Figure~\ref{fig:maps_US} shows the distribution of some of the socio-economic metrics, the population density and COVID-19 prevalence and death rate in all the counties in the USA that we consider in this work. The upper and lower limit of each scale is marked by the 10 percentile and 90 percentile of the distribution respectively. Some clear correlations are visible between the disease spread and the socio-economic metrics. We quantify these correlations later in the work. The COVID-19 prevalence data includes all cases up to the 15$^{th}$ of August, 2020 encompassing both the initial rapid spread in the cities and the later spread in the rural areas. Table~\ref{tab:semetrics} gives an overview of the data that we use. There are a few missing values in the 2018 ACS 5-years data, 7 for Rio Arriba County, NM and 1 for Loving County, TX. We impute those with values from the 2011 census data. We do not include Puerto Rico in this work. Counties where both the number of confirmed cases and the number of deaths are 0 are excluded. The curated data along with the code to perform all our analysis and produce the plots are available at: \href{https://github.com/talismanbrandi/covid-19-USA-SE}{https://github.com/talismanbrandi/covid-19-USA-SE}.

\begin{table}
    \centering
    {\footnotesize
\begin{tabular}{lrrrrrrrrrrr||rr}
\toprule
{} &   Density &   NW &  Poverty &    Income/Cap &  UE &  UI &  Employed &  Labor &  Transit &  MC &  SC &  CCR &  DR \\
\midrule
Mean &    273.14 &  23.50 &    15.60 &  27034.15 &          5.78 &      10.08 &     54.92 &  47.66 &    15.39 &         23.56 &           18.37 &         1067.33 &       28.75 \\
RMS  &   1797.52 &  20.18 &     6.48 &   6512.30 &          2.85 &       5.10 &      8.37 &   7.32 &     6.53 &          5.74 &            4.59 &         1062.69 &       42.17 \\
Min  &      0.04 &   0.00 &     2.30 &  10148.00 &          0.00 &       1.70 &     12.80 &  14.90 &     0.00 &          4.50 &            3.80 &           17.87 &        0.00 \\
25\%  &     16.52 &   7.31 &    11.00 &  22761.75 &          4.00 &       6.20 &     49.40 &  43.52 &    12.02 &         19.70 &           15.42 &          368.81 &        1.74 \\
50\%  &     44.84 &  16.08 &    14.70 &  26243.50 &          5.40 &       9.20 &     55.40 &  48.50 &    14.26 &         23.40 &           18.00 &          739.21 &       13.50 \\
75\%  &    118.55 &  35.24 &    19.10 &  30108.25 &          7.10 &      12.68 &     61.20 &  52.60 &    17.00 &         27.20 &           20.80 &         1431.35 &       37.73 \\
Max  &  71874.13 &  99.27 &    55.10 &  72832.00 &         28.90 &      45.60 &     80.30 &  74.60 &    92.30 &         45.00 &           55.60 &        13984.40 &      425.40 \\
\midrule
\multicolumn{14}{l}{Legend: NW: Non-White, UE: Unemployed, UI: Uninsured, MC: Mean Commute, SC: Senior Citizen, CCR: Confirmed Case Rate, DR: Death Rate.}
\end{tabular}
}
    \caption{\it Statistics for the socio-economic metrics and COVID-19 prevalence data used in this work (excluding counties that have both confirmed case rate and death rate equal to 0). Density is in individuals per sq. mile, Income per capita is in US\$, Mean Commute is in minutes, confirmed case rate and death rate are in units of individuals per 100,000. The rest are in percentages. The explanation of the metrics can be found in the text.}
    \label{tab:semetrics}
\end{table}

\section{Analysis Methods}
\label{sec:analysis}

The primary goal of this study is to examine the importance hierarchy of different factors that contribute to COVID-19 prevalence or death rate. The importance of an independent variable (socio-economic metrics) in determining the dependent variable (COVID-19 prevalence or death rate) can give us insights into how instrumental the independent variable is in shaping the outcome. To achieve this we use machine learning tools and make them interpretable with Shapley values. The procedure we follow can be summarized as:
\begin{itemize}
    \item Examine the correlation between the various socio-economic metrics, population density and the confirmed case rate and death rate keeping focus on only those variables that are significantly correlated with the COVID-19 prevalence and death rates.
    \item Perform a regression using an ensemble of boosted decision trees to build a non-linear model to quantitatively relate the confirmed case rate and the death rate to the socio-economic factors and population density.
    \item Calculate Shapley values to understand variable importance using the model created with the ensemble of boosted decision trees. The Shapley values are used to attribute variable importance.
\end{itemize}
As is clear from this outline, our objective for studying the correlations is to discard evidently unimportant variables. We do not quantitatively use the correlations in our analysis or to infer on variable importance. The reason why we use a machine learning algorithm is because we want to perform a model-agnostic non-linear regression without making any assumptions about the mathematical model in terms of the socio-economic metrics that governs the spread of the disease. Lastly, the reason why we use the Shapley values is to understand the importance of a variable in determining the COVID-19 prevalence and death rates. Now, let us elaborate on these steps.

It is well understood that correlations alone cannot tell the whole story about the importance of a variable in determining the outcome. We use an ensemble of boosted decision trees to perform a regression of the data taking either disease prevalence or death rate as the dependent variable while keeping the socio-economic metrics and population density as the independent variables. This allows us to not assume a functional form for the model but rely on data alone for insights. We emphasize here that we are not trying to build a predictive model using machine learning but rather using it to perform a regression in a model-agnostic manner. We use XGBoost~\cite{10.1145/2939672.2939785}, a scalable end-to-end boosting system for decision trees that is particularly good for sparse data. For data augmentation we use an ensemble of boosted decision trees trained on random selections of the sample with replacement split equally into training and testing sets. This also gives us a stable measure of the accuracy with which we can model the data. We use the coefficient of determination, $R^2$, calculated from the test sample set aside for each boosted decision tree to understand how accurate the regression is. We use the definition
\begin{equation}
    R^2 = 1 - \frac{SS_{res}}{SS_{tot}}
\end{equation}
where $SS_{res}$ is the residual sum of squares computed from the values of the dependent variable in the test dataset and the corresponding prediction by the ensemble of BDTs. $SS_{tot}$ is the total sum of squares computed from the values of the dependent variable in the test dataset and the expectation value of the dependent variable. The value 1 is equivalent to perfect predictions and 0 corresponds to the predictions being equal to the expectation value. The variance in $R^2$ shows how much the weak learners (each BDT) in the ensemble vary in their accuracy of modeling. It is normal for weak learners to differ in their predictions (it is a consequence of bagging within the ensemble). The prediction of an ensemble of weak learners is taken as the average in a regression problem and as a vote in a classification problem in most cases. The accuracy of the ensemble is represented by the mean $R^2$. 

From the stable ensemble we can study the importance hierarchy of the independent variables. This can be done using several methods commonly known as ``feature importance'' in machine learning. We use the SHAP (SHapley Additive exPlanations) values~\cite{NIPS2017_7062}, which is based on Shapley values~\cite{shapley1951notes}, generated by the tree-explainer~\cite{Lundberg:2020vt}. Tree explainers for SHAP values have the added advantage that they work more reliably than other feature importance measures, like Gini or permutation based measures since they are sensitive to correlations between the independent variables which is quite important in the data that we are studying~\cite{2018arXiv180203888L}. It should be noted that we remove the independent variables with small correlations with the dependent variable before we perform the regression with the ensemble of boosted decisions trees hence reducing the chances of unnatural non-zero SHAP values for independent variables that are not important in determining the dependent variable but might be correlated with other independent variables that are important. Hence, we avoid a known problem when using SHAP values for tree-explainers~\cite{2019arXiv190808474S,2019arXiv191013413J}. We will not consider any independent variables that have a correlation of less than 0.1 with the dependent variable. We have checked that this does not adversely affect the goodness of fit of the regression, nor does it change the conclusions about variable importance.

We solve the regression problem $y = f(x_i), i=1,\ldots, n$ with $f$ being the regression model built by an ensemble of boosted decision trees. To determine the SHAP value a game is played with all possible combinations of variables as the players and with the payout being the difference of the predicted value of $y$ from the marginalized value. The SHAP value for the $i^{th}$ variable in determining a value of $y$ is measured from the difference between a value function that includes $x_i$ and the value function which is marginalized over $x_i$. For decision trees this translates into conditional probabilities that are determined by the end-nodes that can be reached in each game. Using the additive property of Shapley values, the importance of the variable $x_i$ is determined by the ensemble average of the absolute SHAP values. The higher the average absolute SHAP value, the more important is the variable in the game of determining $y$. The average SHAP value of any $x_i$ is said to have a positive impact on the model if the SHAP values and the $x_i$ have a positive correlation and a negative impact for negative correlation.

\section{Socio-economic dependencies of COVID-19 prevalence and death rate}
\label{sec:regional}

We will first take a look at the different regions in the USA to get a clear understanding of how the different socio-economic metrics affect the prevalence of COVID-19 in communities with differing demographics. On one hand, studying the variations in the entire nation gives a bigger picture that might be important at the national level. On the other hand, there are large variations in several important factors that can lead to different features being accentuated when different regions are considered separately without these features being lost to aggregation. For example, there are states which have a large average population density and this can drive the spread of infections in a manner different from the ones in which the population density is low. From figure~\ref{fig:maps_US}, it can bee seen that the predominant factor driving the spread of COVID-19 in the states on the east coast is the population density since the disease spread in densely populated urban areas over-shadowed the spread in rural areas in these states. However, this is not the case for the states in southern USA. In addition the states with higher densities were majorly affected in the first wave of the pandemic while the southern states were majorly affected by the second wave. The states on the west coast were affected throughout the extent of the pandemic. There we see that the disease spread has occurred in regions of lower population densities. It is this variation that we want to study in this section by focusing on different clusters of states to glean out the features that are more regionally important. 

In addition we would also like to segment the regions in a way that they hold some commonality in the socio-economic distributions. The northeastern states are have several urban areas that are closely clustered together. However, in the southern states the urban area are spread wide apart and a lot of the population lives in rural areas where the socio-economic conditions are different from the urban areas. On the other hand the states on the west coast are a mix of urban and rural areas with a very different economic structure. While we shall use these criteria for deciding upon how we break up the regions, the method we develop is by no means tied to this and can be used for any other divisions one might want to create. We delineate three different regions for this study:
\begin{itemize}
    \item {\bf High population density regions:} The states with population density over 400 individuals per sq. km: District of Columbia, New Jersey, Rhode Island, Massachusetts, Connecticut, Maryland, Delaware and New York.
    \item {\bf The southern states:} These are the collection of states in the south with the exception of Delaware and District of Columbia which are already included in the previous category. The states included in this category are: Alabama, Arkansas, Florida, Georgia, Kentucky, Louisiana, Mississippi, North Carolina, Oklahoma, South Carolina, Tennessee, Texas, Virginia and West Virginia.
    \item {\bf The west coast:} The three states along the west coast: California, Oregon and Washington are included in this category.
\end{itemize}

\begin{figure}[t!]
    \centering
    \includegraphics[trim=45 20 60 30, clip, width=0.9\textwidth]{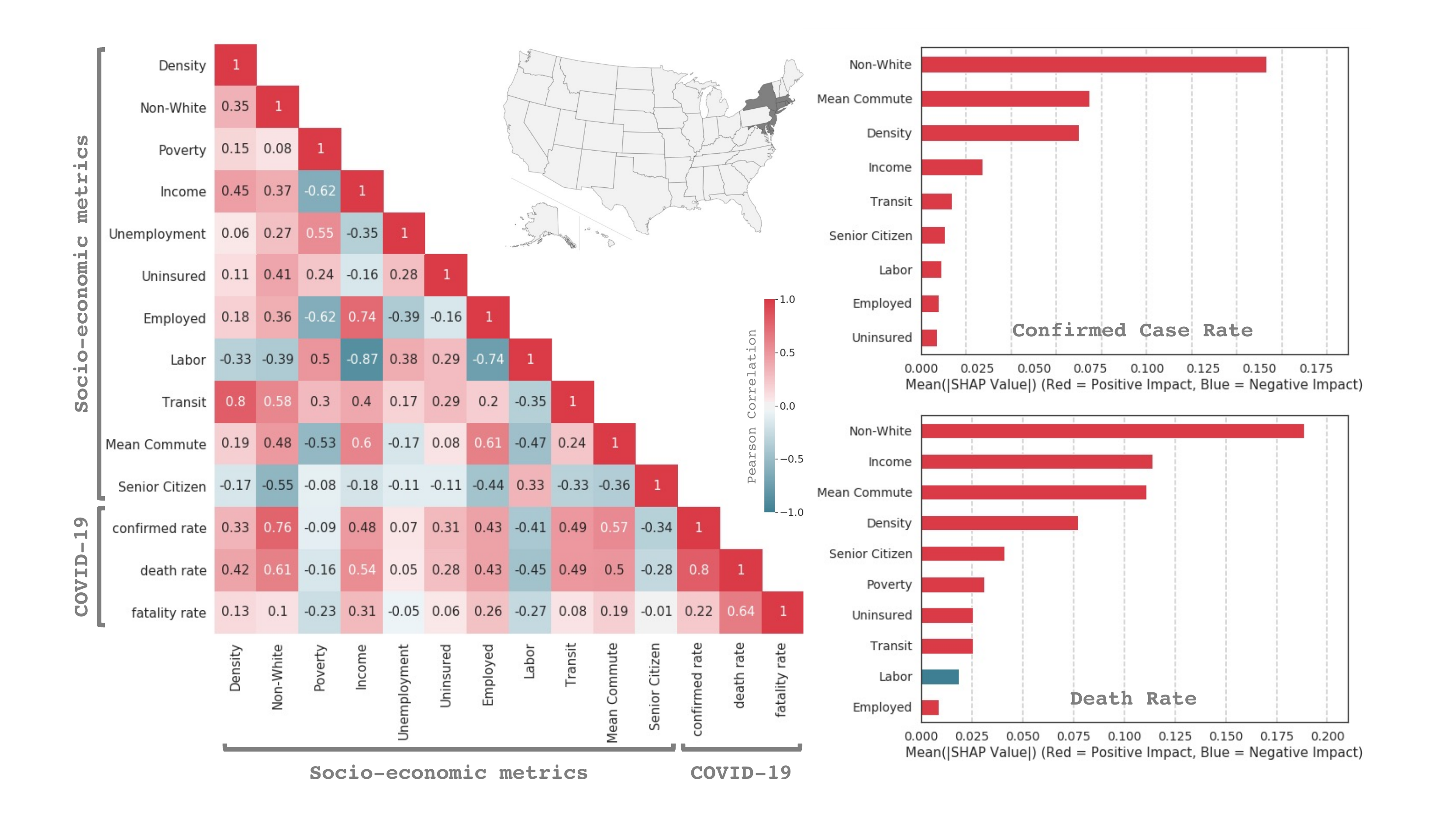}
    \caption{\it Left: The correlation between the socio-economic metrics and COVID-19 confirmed cases rate, death rate, and fatality rate, for the states with highest population density (> 400 per square mile): District of Columbia, New Jersey, Rhode Island, Massachusetts, Connecticut, Maryland, Delaware and New York. Right: Averaged absolute SHAP values of the different contributing variables. Higher SHAP values indicate greater importance. Shapley values are expressed in $\log(x)$ where $x$ is the confirmed cases per 100,000 and deaths per 100,000 in the upper and lower right plots respectively.}
    \label{fig:corr_EC}
\end{figure}
\begin{figure}[h!]
    \centering
    \includegraphics[trim=45 20 60 30, clip, width=0.9\textwidth]{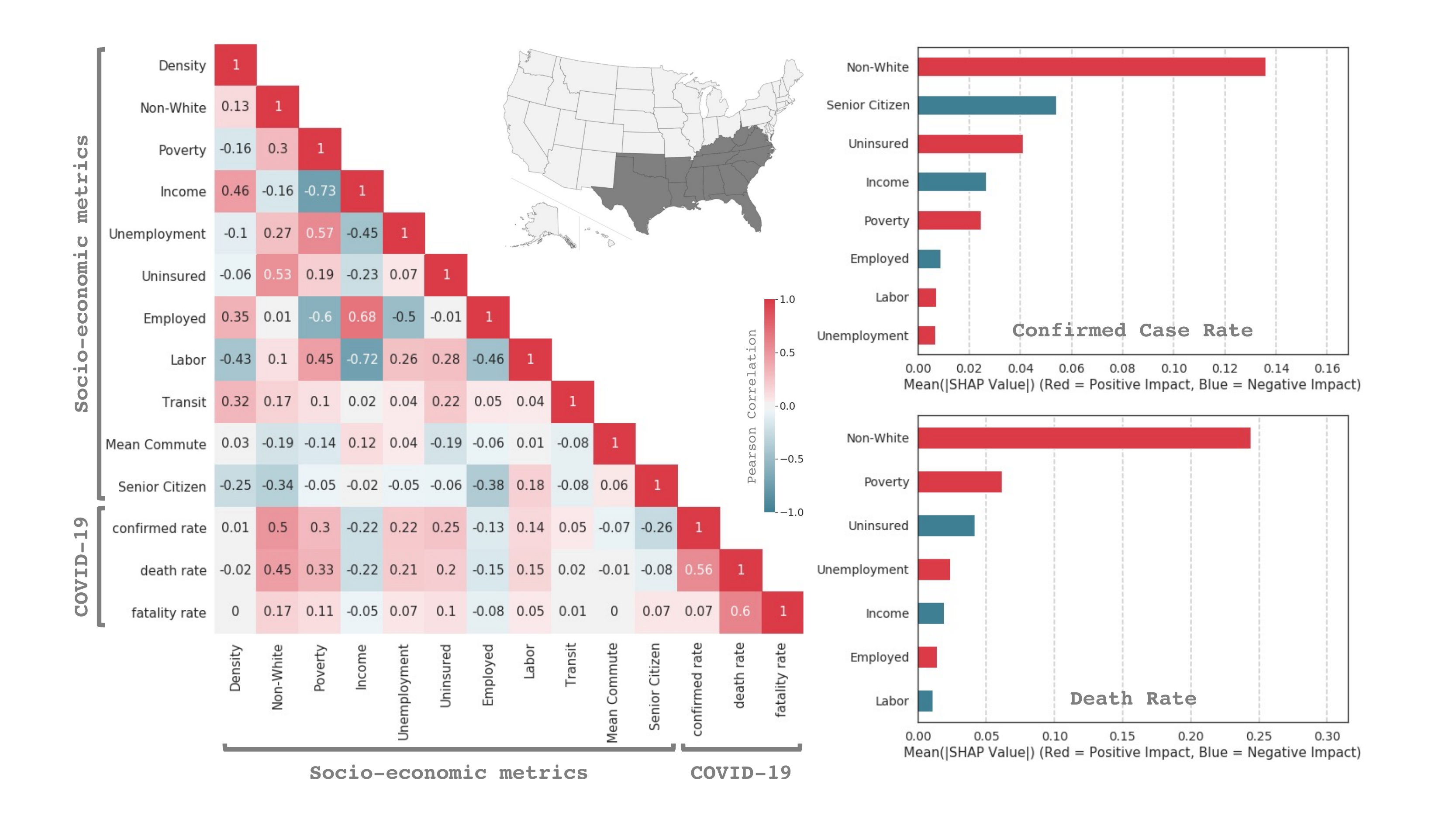}
    \caption{\it Left: The correlation between the socio-economic metrics and COVID-19 confirmed cases rate, death rate, and fatality rate, for the southern states: Alabama, Arkansas, Florida, Georgia, Kentucky, Louisiana, Mississippi, North Carolina, Oklahoma, South Carolina, Tennessee, Texas, Virginia and West Virginia. Right: Averaged absolute SHAP values of the different contributing variables. Higher SHAP values indicate greater importance. Shapley values are expressed in $\log(x)$ where $x$ is the confirmed cases per 100,000 and deaths per 100,000 in the upper and lower right plots respectively.}
    \label{fig:corr_SS}
\end{figure}

\subsection{High population density regions}
\label{sec:EC}

Figure~\ref{fig:corr_EC} depicts how COVID-19 prevalence and death rate are correlated with  socio-economic metrics for the states that have a very high population density. As can be expected, both the confirmed case rate and the death rate are positively correlated with population density. Besides this, metrics that point towards higher mobility and mixing amongst the population like Income, Employed, Transit and Mean Commute show large positive correlations with both the confirmed case rate and the death rate. What is striking is the large and dominating positive correlations of the confirmed case rate and death rate with the non-white fraction of the population. Also notable is the negative correlation of the confirmed case rate and the death rate with fraction of Senior Citizen in a county. This implies that the spread of COVID-19 is higher in communities with a larger fraction of younger individuals although clinical evidence shows that older people are more at risk of infection and mortality.

The modeling accuracy from the ensemble of boosted decision trees is 72\% $\pm$ 5\% for the confirmed case rate and 30\% $\pm$ 13\% for the death rate implying that the ensembles are quite effective in modeling the confirmed case rate but not so effective in modeling the death rate. The plots on the right side of figure~~\ref{fig:corr_EC} display the importance of each metric in determining COVID-19 prevalence and death rate. It shows that the most important feature in determining the spread of COVID-19 is the non-white fraction of the population while the second most important one is Mean Commute followed by population density for confirmed case rate and Income for death rate which is almost as important as Mean Commute. The SHAP values for the top three metrics in both the plots are quite close showing that just one metric does not dominate the determination of the disease prevalence and death rate but a combination of the first few is important. All of these put together point to the conclusion that counties with a larger fraction of racial and ethnic minorities are the worst affected by the pandemic in densely populated states.

\subsection{The southern states}
\label{sec:SS}

The pattern seen in Figure~\ref{fig:corr_SS} is a bit different. Here we see that the disease spread is highly correlated with the non-white fraction of the population in a county and is not so correlated with any measures of mobility or economic conditions. The disease spread completely uncorrelated with population density pointing to the fact that the disease is spreading independent of the population density of the county. The correlation matrix also shows that while the confirmed case rate are relatively lower in counties with higher fraction of senior citizen, the death rate is not correlated with this. The SHAP values on the right side of the plot corroborate what the correlation matrix suggests. The single most important metric that determines the spread of COVID-19 and deaths due to the disease is the non-white fraction of the population in the county. This is also visually evident from figure~\ref{fig:maps_US} when comparing the top left panel with the bottom panels. The modeling accuracy of the confirmed case rate by the ensemble of boosted decision trees is 44\% $\pm$ 2\% and for the death rate it is 28\% $\pm$ 2\%. Both the ensembles show significant stability in their accuracy but the confirmed case rate is modeled more accurately than the death rate.

\begin{figure}[t!]
    \centering
    \includegraphics[trim=45 20 60 30, clip, width=0.9\textwidth]{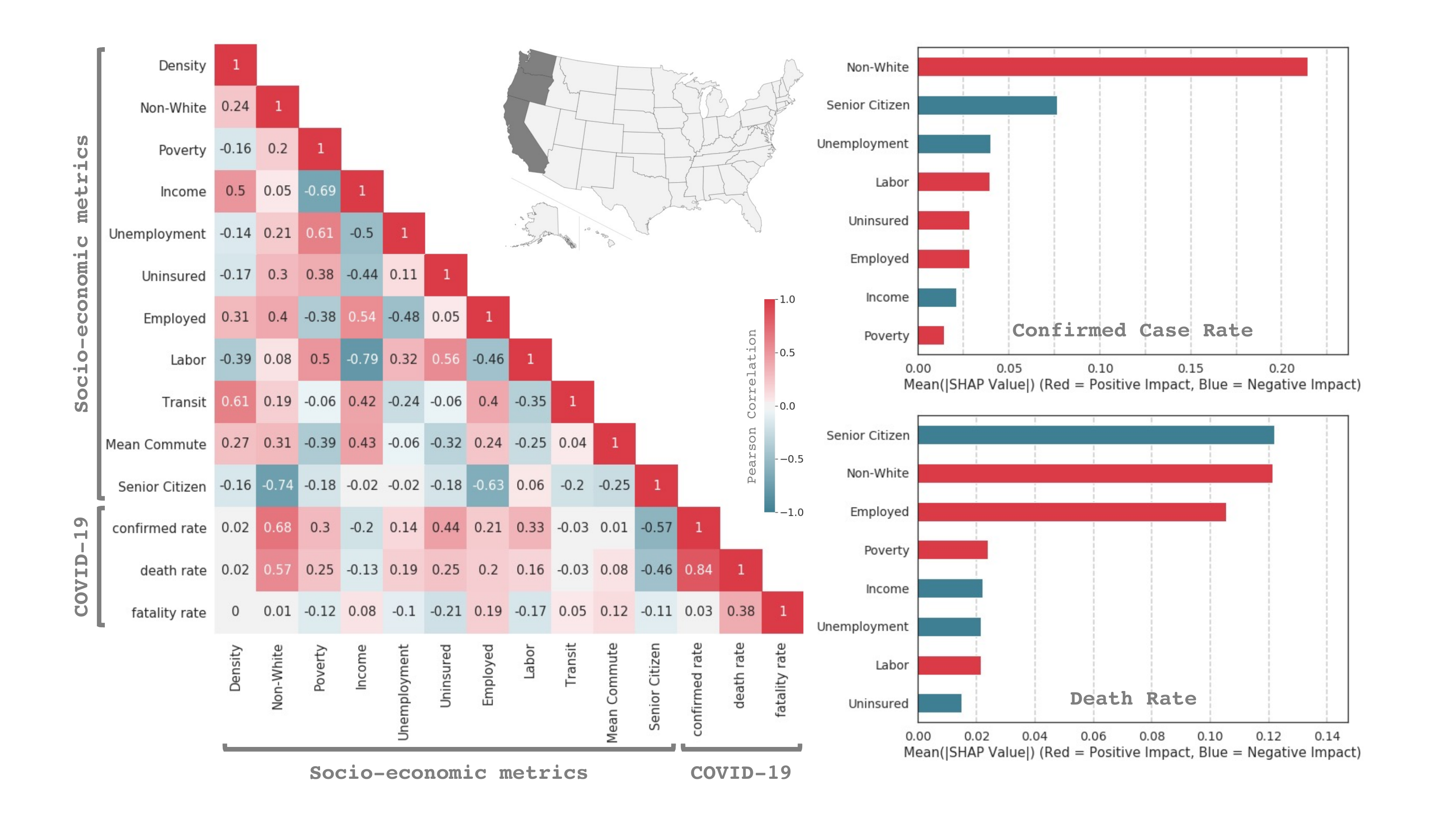}
    \caption{\it Left: The correlation between the socio-economic metrics and COVID-19 confirmed cases rate, death rate, and fatality rate, for the west coast states: California, Oregon and Washington. Right: Averaged absolute SHAP values of the different contributing variables. Higher SHAP values indicate greater importance. Shapley values are expressed in $\log(x)$ where $x$ is the confirmed cases per 100,000 and deaths per 100,000 in the upper and lower right plots respectively.}
    \label{fig:corr_WC}
\end{figure}
\begin{figure}[h!]
    \centering
    \includegraphics[trim=45 20 60 30, clip, width=0.9\textwidth]{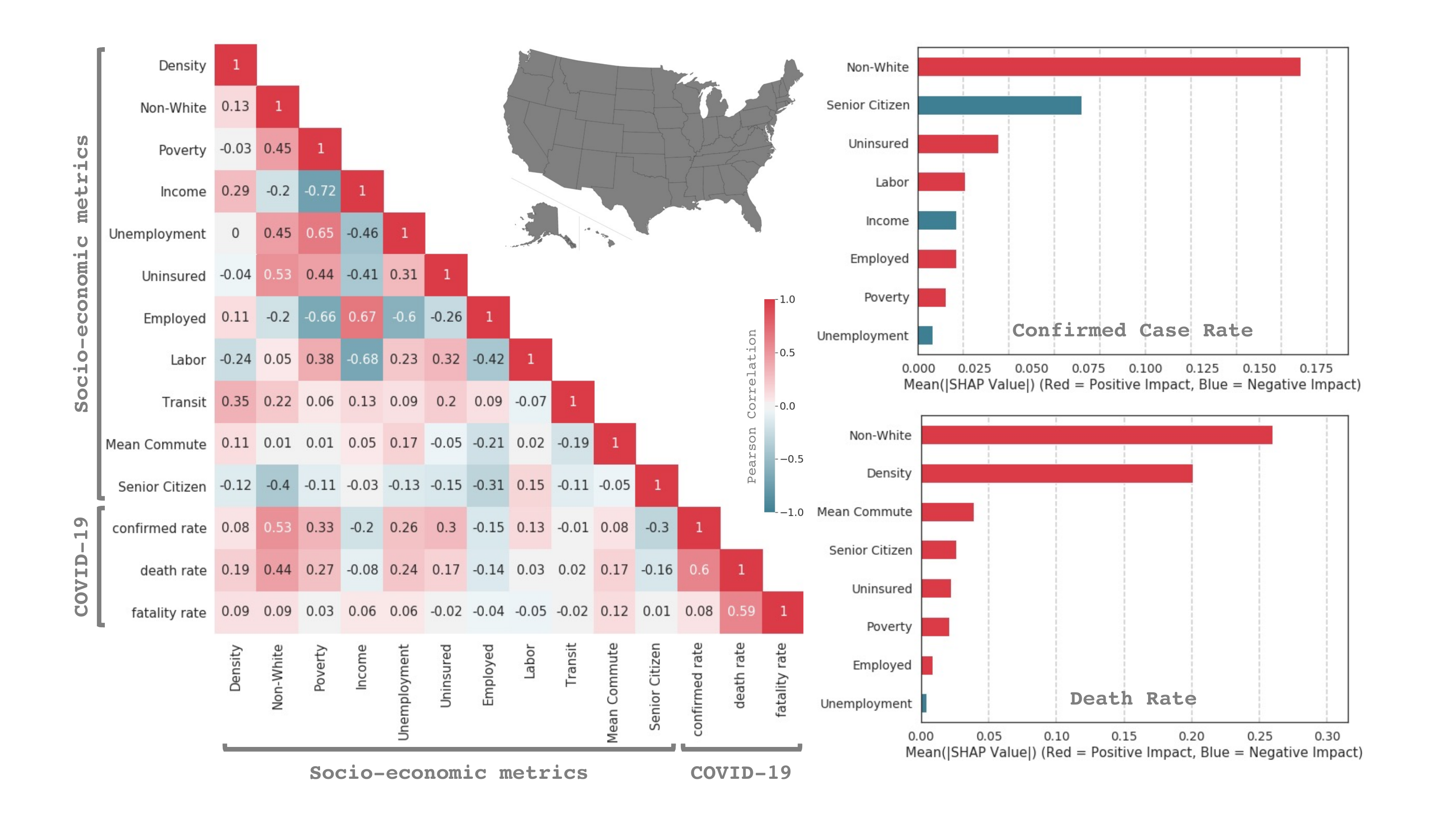}
    \caption{\it Left: The correlation between the socio-economic metrics and COVID-19 confirmed cases rate, death rate, and fatality rate, for the USA. Right: Averaged absolute SHAP values of the different contributing variables. Higher SHAP values indicate greater importance. Shapley values are expressed in $\log(x)$ where $x$ is the confirmed cases per 100,000 and deaths per 100,000 in the upper and lower right plots respectively.}
    \label{fig:corr_USA}
\end{figure}
\subsection{The west coast}
\label{sec:WC}

The states along the west coast show a distinct pattern. From the left panel of figure~\ref{fig:corr_WC} we see that the disease spread has a large negative correlation with the fraction of senior citizen in a county. This can be understood from the fact that counties with larger fraction of younger individuals tend to move and mingle a lot more driving up the disease prevalence. This can be better understood from the fact that the fraction of the population that is employed is also negatively correlated with the fraction of senior citizen in a county. In addition, counties with larger fraction of senior citizen also tend to have a smaller non-white fraction of the population. 

The modeling accuracy for the confirmed case rate is 58\% $\pm$ 8\% while for the death rate it is 23\% $\pm$ 12\%. In the plots on the right side for the SHAP values, in agreement with the correlation matrix, we see that the determining factor for the disease spread and deaths due to COVID-19 is the fraction of senior citizen and the non-white fraction of the population in a county. The former leaves a negative impact while the latter leaves a positive impact. This implies that higher fractions of senior citizen in a county are correlated with lower SHAP values which in turn means that the fraction of senior citizen tend to lower COVID-19 prevalence and deaths. While this is quite counter-intuitive since COVID-19 affects the aged preferentially, one must understand that these numbers point at how the disease spreads in a community and not to individuals. It really highlights the roles of these metrics at a community level and accentuates the necessity to limit movement and mingling within a community if the disease spread has to be mitigated.

\subsection{USA as a nation}
\label{sec:US}

Finally, we take a look at USA as a whole nation. Of course, the regional features will be lost to aggregation since they are very different in different regions. However, the correlation matrix in figure~\ref{fig:corr_USA} clearly points to the non-white fraction of the population as being the most correlated with COVID-19 confirmed case rate and death rate. Secondly, the confirmed case rates are also negatively correlated with the fraction of senior citizen in a county pointing to the fact that the disease is preferentially spreading in counties with younger population on an average.

In agreement, the SHAP value plots on the right determine that the the non-white fraction of the population and the fraction of senior citizen in a county are the most important determining factor for the propagation of the disease. For the death rate the non-white fraction of the population and the population density are the most important factors. The fact that population density is uncorrelated with the confirmed case rate while being correlated with the death rate is an artifact of the way the disease spread in the USA. In the first wave when the disease spread in the most densely populated states, the death rate was much higher as we seen before. However, the second wave has seen a surge in the number of cases in less densely populated areas of the southern states and the west coast states but the death rate have been relatively low. This can possibly be because a healthcare system is more likely to be overwhelmed in a more densely populated area. The modeling accuracy for the confirmed case rate is 42\% $\pm$ 2\% while for the death rate it is 41\% $\pm$ 2\%.

\section{Summary}
\label{sec:summary}

The COVID-19 pandemic has brought about extensive data collection during a wide-spread pandemic at unprecedented scales. In our work, we use this data to study trends that are not obvious to us without the intervention of hard numbers and studies of complex correlations. To achieve our goal, we study several socio-economic metrics and show that, while predictive modeling is not possible, the study of trends can lead to prescriptive conclusions with which first order tuning to public and healthcare policies can be made. Deconvoluting causations will take a lot more work which we leave for the future. In this sense our work should be taken with the correct set of caveats and with an understanding that, at best, our conclusions stand in aggregations at the county level and not at the level of effects on individuals. 

It is known that superspreading events are not uncommon in the transmission of COVID-19~\cite{Frieden2020,Streeck2020}. In our analysis we do not take into account possible superspreading events explicitly. While such events typically happen in regions with higher population density, this cannot be concluded upon without an extensive study. However, since not many superspreading events have been documented in the USA, we assume that our analysis is not significantly biased by our lack of quantification of the effects of superspreading. We have not taken into account the effects of limited testing at the beginning of the pandemic. The primary conclusions from our study are:
\begin{itemize}
\setlength\itemsep{0em} 
    \item The spread of COVID-19 can be an "urban phenomenon" when the population density is very high. In this regard, large dense cities require special attention for the mitigation of the spread.
    \item In areas of lower population densities such as suburban and rural areas, the spreading of the disease is not governed by the population density of the region.
    \item It can be clearly seen from the data that regions with larger fraction of racial and ethnic minorities are being affected the most, a conclusion supported by several other studies~\cite{MILLETT2020,10.1001/jama.2020.6548,doi:10.1056/NEJMp2021971,10.15585/mmwr.mm6933e1,10.1001/jama.2020.11374,DIMAGGIO20207,MILLETT202037,doi:10.1080/13557858.2020.1853067,Pareek2020,Laurencin2020}. This is particularly true for the southern states while being less relevant for the west coast states.
    \item While COVID-19 preferentially affects the elderly at an individual level, the spread of the disease is dependent on human mobility within a region. This leads to an anti-correlation between the age of a population in a county and how easily the disease spreads there as seen in both the southern states and in the west coast states. In other words, while the elderly are most affected by COVID-19, it spreads more easily in regions where the population is relatively younger.
    \item Fatality rates are mostly uncorrelated with the socio-economic metrics and population density. This implies that death due to COVID-19 amongst those who are already infected does not depend on the socio-economic classes.
    \item The study including all the states considered in this work fails to highlight the different drivers in different regions. This clearly shows that the spreading characteristics of COVID-19 does not show representative trends at the national level and studies must be made at regional levels.
    \item Our results should be taken as descriptive or, at best, prescriptive and by no means predictive about the spread of COVID-19. For the confirmed case rate for the states with high population density, the accuracy of the model is quite high and falls in the predictive regime validating the robustness of the method.
\end{itemize}

We would like to draw the attention of the readers to certain subtleties in the data that we were not able to fully address in this work. The data on confirmed case rates and death rates were collected at the early stage of the pandemic. For the confirmed case rate, initially, insufficient testing affected the data. In this work we implicitly assume that there were no testing biases that would be correlated with the socio-economic metrics that we have considered. This can possibly lead to certain biases in this analysis since those that had a higher risk of getting severely infected or had shown more prominent symptoms might have been preferentially tested and this is known to be correlated with age. However, death rates are less affected by any implicit biases due to such correlations since they are better recorded and are not affected by variables such as testing strategies.

There are several directions in which this work can be extended. One of the important directions is a comprehensive study of how COVID-19 prevalence and death rate are correlated with comorbidities at the county level. Underlying heath conditions are correlated with several socio-economic metrics, many of which we consider here. Detailed studies of the correlation between underlying health conditions and COVID-19 at the county level has been recently completed ~\cite{10.15585/mmwr.mm6929a1}. An analysis of chronic conditions amongst recipients of medicare fee-for-service~\cite{doi:10.1089/pop.2019.0231} can also be included to understand the correlation with the spread of COVID-19 and its relation to a slice of the population. It would also be insightful to connect data on inter-county mobility to understand how counties can be shielded from one another during a pandemic. We leave these as future work.

\section*{Acknowledgements}
This research was supported in part through the Maxwell computational resources operated at DESY, Hamburg, Germany. We have gained much from insightful discussions with Matteo Rinaldi, Tannista Banerjee, Hyunju Kim, Aditi Sengupta and Christian Suharlim.

\vspace{0.25cm}\noindent\textbf{Disclaimer:}
This work solely reflects the views of the authors in their academic capacity and does not represent the views of any company or client.

\bibliography{bibliography}




\end{document}